\documentclass[10pt,a4paper]{article}
\usepackage{amssymb,amsbsy,graphicx,subfigure,times}
\usepackage{hyperref}
\usepackage[latin1]{inputenc}
\usepackage{amsmath,amsfonts,amssymb,graphics,graphicx,epsfig,color}
\usepackage{subfigure}
\usepackage{fancyhdr,makeidx}
\usepackage{color}

\newcommand{\N}{\cal N}

\newcommand{\tr}{{\rm Tr}\,}

\newcommand{\gr}[1]{\boldsymbol{#1}}
\newcommand{\be}{\begin{equation}}
\newcommand{\ee}{\end{equation}}
\newcommand{\bea}{\begin{eqnarray}}
\newcommand{\eea}{\end{eqnarray}}

\newcommand{\sig}{\gr{\sigma}}

\begin{document}
\title{Generation of continuous variable squeezing and entanglement
of trapped ions in time-varying potentials}
\author{
Alessio Serafini,$^{1}$ Alex Retzker,$^{2,3}$ and Martin B.~Plenio$^{2,3}$\\
\small $^{1}$ Department of Physics \& Astronomy, University College London,\\
\small Gower Street, London WC1E 6BT, UK\\
\small $^{2}$ Institute for Mathematical Sciences, 53 Prince's Gate,\\
\small Imperial College London, London SW7 2PG, UK\\
\small $^{3}$ QOLS, Blackett Laboratory, Imperial College London,\\
\small London SW7 2BW, UK
}
\date{15-04-2009}

\maketitle

\begin{abstract}
We investigate the generation of squeezing and entanglement
for the motional degrees of freedom of ions in linear traps, 
confined by time-varying and oscillating potentials, comprised 
of an DC and an AC component. We show that high degrees of 
squeezing and entanglement can be obtained by controlling 
either the DC or the AC trapping component (or both), and by 
exploiting transient dynamics in regions where the ions' motion 
is unstable, without any added optical control. 
Furthermore, we investigate the time-scales over which the
potentials should be switched in order for the manipulations 
to be most effective.
\end{abstract}


\section{Introduction and scope of the paper}\label{intro}
The manipulation of continuous variable quantum information is 
by now a well established area of research \cite{martinrev},
that has led to a number of remarkable experimental 
and technological advances \cite{braunstein05}. So far, the 
physical system of choice in the arena of continuous variables
has certainly been the electromagnetic field, mostly because of
the ease with which it can be coherently manipulated and distributed 
in space, making it exceptionally suitable for quantum communication 
tasks.
Yet, notwithstanding the clear benefits
and successes of quantum optical degrees of freedom,
there are reasons why one would be interested in exploring alternatives.

Firstly, while the entanglement between two continuous variable
degrees of freedom could be, in principle, unbounded, the degrees
of entanglement achievable in practice for quantum optical systems
are severely limited by the photons' reluctance to interact with each other.
Entanglement (and squeezing) are obtained in continuous variable systems
through interactions mediated by parametric crystals:
to the best of our knowledge, assuming
the highest reported degrees of squeezing (corresponding to a noise
reduced to $0.1$ vacuum units) \cite{takeno07,vahlbruch08}
and perfect mixing operations, one can achieve 
a logarithmic negativity $E_{\N} \simeq 3 \,{\rm ebits}$
(see also \cite{french} for a ``measured'', {\em i.e.}~inferred from state
reconstruction, value of $1.6 \,{\rm ebits}$).

Secondly, while extremely good at traveling,
electromagnetic fields don't make such good static degrees of 
freedom: even though they can be trapped in cavities,
one is often confronted with a challenging and impractical 
trade-off between keeping the cavity open to external fields 
in order to access the quantum information and isolating the 
cavity to reduce losses and thus decoherence. This problem 
could be partially solved by mapping the quantum state of 
light into polarised atomic clouds memories \cite{Polzik1,Polzik2}.
Though such a technology has been pioneered and successfully tested,
its performance is still far from ideal, whence it may be desirable to
resort to other static degrees of freedom
allowing for the direct manipulation of continuous variable
quantum information.

An extremely promising candidate to this aim, which might
potentially address both the aforementioned issues, is
represented by the motional degrees of freedom of trapped ions.
The control of positions and momenta of trapped atoms and ions has been
successfully implemented in several past experiments, 
both for its own sake \cite{heizen1990,meekhof1996,vogel1995,poyatos1996a,bardroff1996} 
and to address internal degrees of freedom
({\em e.g.}~in realising prototypes of the Cirac-Zoller ionic quantum computer
\cite{heizen1990,meekhof1996,vogel1995,poyatos1996a,bardroff1996}).
In the case of ions,
Coulomb interactions between the ions could be exploited
to generate entanglement between motional degrees of freedom,
while the long achievable trapping times would account for the need of good
static degrees of freedom.

Indeed, the quantum properties of mechanical degrees of freedom 
have gained a large amount of interest recently, either in 
terms of squeezing or in terms of entanglement properties. 
Most of such  investigations have focused around nannomechanical and 
micromechanical oscillators \cite{nanoexp,EisertPlenioBoseHartley2004,
HartmannPlenio2008}
but, recently the entanglement of 
motional degrees of ions was created and measured as well \cite{Jost09}.
 
Here we shall focus on the transverse (``radial'') motions, 
which can be individually addressed, and where phonons can be locally defined, 
thanks to the tightness of the transverse confinement 
\cite{zhu2006,nonlinear}. 
In a previous contribution from the authors \cite{citeus}, 
it was shown that comprehensive manipulations of such radial degrees 
of freedom could be realised for two and more ions in a linear Paul 
trap by controlling the radial trapping frequencies. More 
specifically, it was shown that the capability to control each 
individual trapping potential in the array would allow for the 
implementation of any linear operation on the motions, including 
squeezing. Moreover, creation of high degrees of bipartite and 
multipartite entanglement was shown to be possible with
only global control of the trap potential.
The transmission of quantum information through the chain of ions,
in both qubit and genuine continuous variable form, was also studied
and shown to be achievable.
Besides, it was indicated how multipartite entanglement of three ions
could be put to use to violate Bell-like inequalities and demonstrate quantum
non-locality.

In all these coherent manipulations, the 
only kind of experimental control supposed was 
the possibility of tuning and changing the electric trapping potentials:
no optical control through laser pulses was required.
However these theoretical findings, promising as they are, were all
derived assuming two major idealisations:
\begin{itemize}
\item{the changes of the trapping potential,
which are the main way to manipulate the quantum states, 
were assumed to be instantaneous;}
\item{the potentials were assumed to be static in time,
which is only approximately true in a Paul trap, if the static component of the
trapping field is large
with respect to the amplitude of the oscillating component.}
\end{itemize}
In this note, we will relax these two assumptions and study
how the dynamics of the radial modes is affected if
finite switching times and oscillating trapping potentials
are taken into account.

The paper is organised as follows.
In section \ref{hamil} we describe the
time-dependent Hamiltonian governing the evolution
of the system and define the experimental parameters
under control.
In section \ref{squeezzo} we address the generation of squeezing
in the position and momentum of a single trapped ion,
while
in section \ref{entanglo} we will present results on the generation of entanglement
of two trapped ions, considering
the effect of finite switching times and of oscillating potentials.
Notice that continuous variable squeezing and entanglement are closely
related as, essentially, entanglement manifests itself in the squeezing of
combined quadratures, like in the Einstein Podolski Rosen seminal example.
Finally, some concluding words and future perspectives
are given in section \ref{outro}.

\section{The trapping potential}\label{hamil}
We shall consider radial modes (along a transverse direction
with respect to the trap's axis)
of one or two ions of mass $m$ and charge $ze$ in a linear Paul trap.
Let $\hat{X}$ and $\hat{P}$ be the
position and momentum operators of a single ion
associated to the considered radial degree
of freedom, then the Hamiltonian governing the dynamics
of $\hat{X}$ and $\hat{P}$ in the quadrupole trapping field is
\be
\hat{H}(a,q) = \frac{\hat{P}^2}{2m} + \frac{m\Omega^2}{8}
\left( a + 2q\cos(\Omega t) \right) \hat{X}^2 \; ,  \label{single}
\ee
where $\Omega$ is the frequency of the oscillating trapping potential,
while $a$ and $q$ are dimensionless parameters which determine,
respectively, the amplitude of the DC trapping field and of the AC
oscillating trapping field. The parameters and factors are chosen so that
the resulting $a$ and $q$ are the same as in the classical Mathieu equation,
whose solutions discriminate between stable (generally trapped) and
unstable motions of the ions in the trap \cite{ghosh}.


In the case of two ions, the Hamilonian contains also
an interaction term due to the Coulomb repulsion between the ions:
we shall approximate this term to the second order in the
displacement, thus obtaining a quadratic term coupling the
oscillations of the two ions.  
Notice that this ``harmonic'' approximation is very accurate 
in our case, where 
transverse to longitudinal potential ratios will be larger than $0.1$.
Under such conditions, the ratio between radial displacements
and distance between neighbouring ions is at most about $0.02$.
Hence, fourth and higher order terms in the displacements
are at least $(0.02)^2 \simeq 4\times10^{-4}$
times smaller than the considered second order terms
and can be safely neglected. 
However, we should note that this would not be the case anymore for 
very large amounts of radial squeezing: in this case 
the anharmonic corrections would have to be taken into account. 
In the present study we shall restrict to cases where the squeezing, while large,
is still small enough for the harmonic approximation to hold.
The Hamiltonian $H_2$ for two ions read
\bea
\hat{H_2}(a,q) &=& \frac{\hat{P}_1^2+\hat{P}_2^2}{2m}
+ \frac{m\Omega^2}{2}
\left[\frac14\left(a + 2q\cos(\Omega t)\right) - \xi \frac{\Omega_{L}^2}{\Omega^2} \right] \nonumber\\
&&\left(\hat{X}_1^2+\hat{X}_2^2\right) + 2\xi \Omega_L^2 \hat{X}_1\hat{X}_2 \; . \label{couple}
\eea
The factor $\xi$ comes from the Coulomb interaction at lowest order and
is actually equal to ${z^2e^2/(4\pi\varepsilon_{0}m\Omega_L^2d^3)}\simeq0.5$, where
$d$ is the distance between the ions and
$\varepsilon_0$ is the vacuum electric permittivity \cite{james98}.
Notice that $\xi$ itself does not depend on the mass of the ions nor on the longitudinal trapping
frequency $\Omega_L$, although the Hamiltonian terms in which it enters can be tuned
by adjusting $\Omega_L$.

In this study, we will consider situations where the parameters $a$ and $q$
can be controlled and changed over time by a hypothetical experimentalist.
However, we will not assume such changes to occur instantaneously.
The state of the system will thus evolve under the time-varying
Hamiltonians $\hat{H}(a(t),q(t))$ and $\hat{H}_2(a(t),q(t))$.

As initial states, we will assume the ground state of the Hamiltonian $\hat{H}_{pw}$
with trapping frequency $\omega_{pw} = \Omega^2\left(a(0)+q(0)^2/2\right)/4$:
\be
\hat{H}_{pw} = \frac{\hat{P}^2}{2m} + \frac{m}{2}\omega_{pw}^2 \hat{X}^2 \; . \label{pwell}
\ee
This is the initial effective trapping frequency in the so-called ``potential well
model'' \cite{ghash}, when the positions of the ions in the trap can be separated into
a comparatively small and fast micro-motion, and a comparatively
large and slow dominant term.
Note that this initial state is Gaussian
({\em i.e.}, it has Gaussian Wigner and characteristic functions
and is hence completely characterised by the first and second statistical moments
of positions and momenta), while the subsequent
dynamics is linear and thus preserves the Gaussian character of the state.
Therefore the dynamics can, under such conditions, be integrated
numerically with straightforward techniques.
To this aim,
we employed the Runge-Kutta (RK4) method and cross checked it
against the piece-wise exponentiation of the Hamiltonian matrix:
the two methods yielded essentially coincident results for small enough 
time-steps (actually, at variance with RK4, 
the piece-wise exponentiation carries a small second order error, which would 
however be hardly noticeable in the results we will present).

As mentioned above a Gaussian state $\varrho$ is completely determined
by its first and second moments: first moments will not be of any concern here,
as they can be unitarily adjusted and do not affect the quantities we set to study.
The second moments can be conveniently grouped together in
the ``covariance matrix'' (CM) ${\gr\sigma}$, with entries
${\sigma}_{jk} \equiv \tr{[\{\hat R_j , \hat R_k\} \varrho]}/2
-\tr{[\hat R_j \varrho]}\tr{[\hat R_k \varrho]}$,
in terms of the vector of canonical operators:
$\hat{R} = \left( \hat{X},\hat{P}\right)$ for one ion and 
$\hat{R} = \left( \hat{X}_1,\hat{X}_2,\hat{P}_1,\hat{P}_2 \right)$
for two ions \cite{martinrev,martinshash,mga}.

\begin{figure}[t!]
\begin{center}
\includegraphics[scale=0.4]{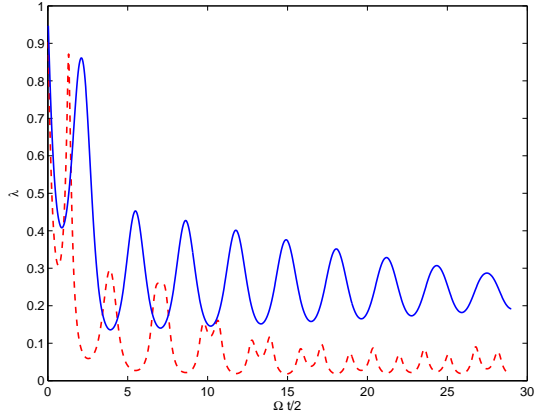}
\caption{Squeezing (smallest eigenvalue of the CM) for an initial state $\sig_0$
(ground state of $\hat{H}_{pw}$) evolving
under the Hamiltonian (\ref{single}) for varying $a$'s and $q$'s.
Red (dashed) curve: $a(t)=-0.001+(-0.1+0.001)\Omega t/4$ for $0\le t \le 4/\Omega$,
$a(t) = -0.1 + (-0.001+0.1)(\Omega t/4-1)$ for $4/\Omega \le t \le 8/\Omega$,
$a(t)=-0.001$ for $t\ge 8/\Omega$, and $q(t)=0.1$ $\,\forall t$.
Blue (continuous) curve: $a(t)=0.0001+(0.01+0.0001)\Omega t/4$ for $0\le t \le 4/\Omega$,
$a(t) = 0.01 + (0.0001-0.01)(\Omega t/4-1)$ for $4/\Omega \le t \le 8/\Omega$,
$a(t)=0.0001$ for $t\ge 8/\Omega$, and $q(t)=0.01$ $\,\forall t$.\label{Sq1}}
\end{center}
\end{figure}
\section{Generation of squeezing}\label{squeezzo}

In order to study the generation of squeezing by time-varying potentials
we will consider a single ion, starting from the
ground state of the Hamiltonian (\ref{pwell}), and evolving in time under
the Hamiltonian (\ref{single}) for properly chosen $a(t)$ and $q(t)$.
However, we will consider the rescaled quadratures $\hat{x}=\sqrt{m\omega_{pw}}\hat{X}$
and $\hat{p}=\hat{P}/\sqrt{m\omega_{pw}}$, so that this ground state $\sig_0$,
which will constitute our reference for the vacuum, reduces
to a Gaussian state with covariance matrix equal to the identity (in our units).
Once the dynamics is solved and the CM $\sig_t$ at subsequent time is obtained,
it will thus suffice to evaluate the smallest eigenvalue $\lambda$ of $\sig_t$ as
a signature of squeezing: the smallest $\lambda$
(compared to $1$) the largest the squeezing.

Fig.~\ref{Sq1} shows two cases with small $a$'s and $q$'s, and with $|a(0)|\ll |q(0)|$,
where the potential well model is very accurate.
For the red (dashed) curve, $q(t)=0.1$ at all times while $a$ starts from $a(0)=-0.001$ and
then switches, linearly over a time interval $4/\Omega$, to $a(4/\Omega)=-0.01$ to finally
decrease back to the initial value $-0.001$. The system evolves then under the original Hamiltonian from
$t=8/\Omega$ on. The static DC field is always repulsive in this case, but the ion's
stability is guaranteed by the AC component. Besides, even though the initial and final conditions are stable,
the ion briefly goes through a region of instability through this motion.
As can be seen, if such passages can be carried out quickly enough
not to lose the ion in the unstable region (that is, on the time-scale of $\Omega^{-1}$),
then the average squeezing resulting from the change of potential can be remarkably high,
even up to $0.1$ vacuum units like here.
The blue (continuous) curve shows instead the case $q(t)=0.01$ at all times and $a$
increasing from $0.0001$ to $0.01$ for a time $0<t<4/\Omega$, and then
decreasing back to $0.0001$ for $4/\Omega<t<8/\Omega$. In this case as well
a relatively small, but rapid, change in the DC field yields a considerable degree of squeezing.
However these values of squeezing, around $\lambda=0.25$, are well below the previous instance,
when the system underwent unstable regimes (whereas in this case the ion's motion is stable for
all $a$'s). 

\begin{figure}[t!]
\begin{center}
\includegraphics[scale=0.3]{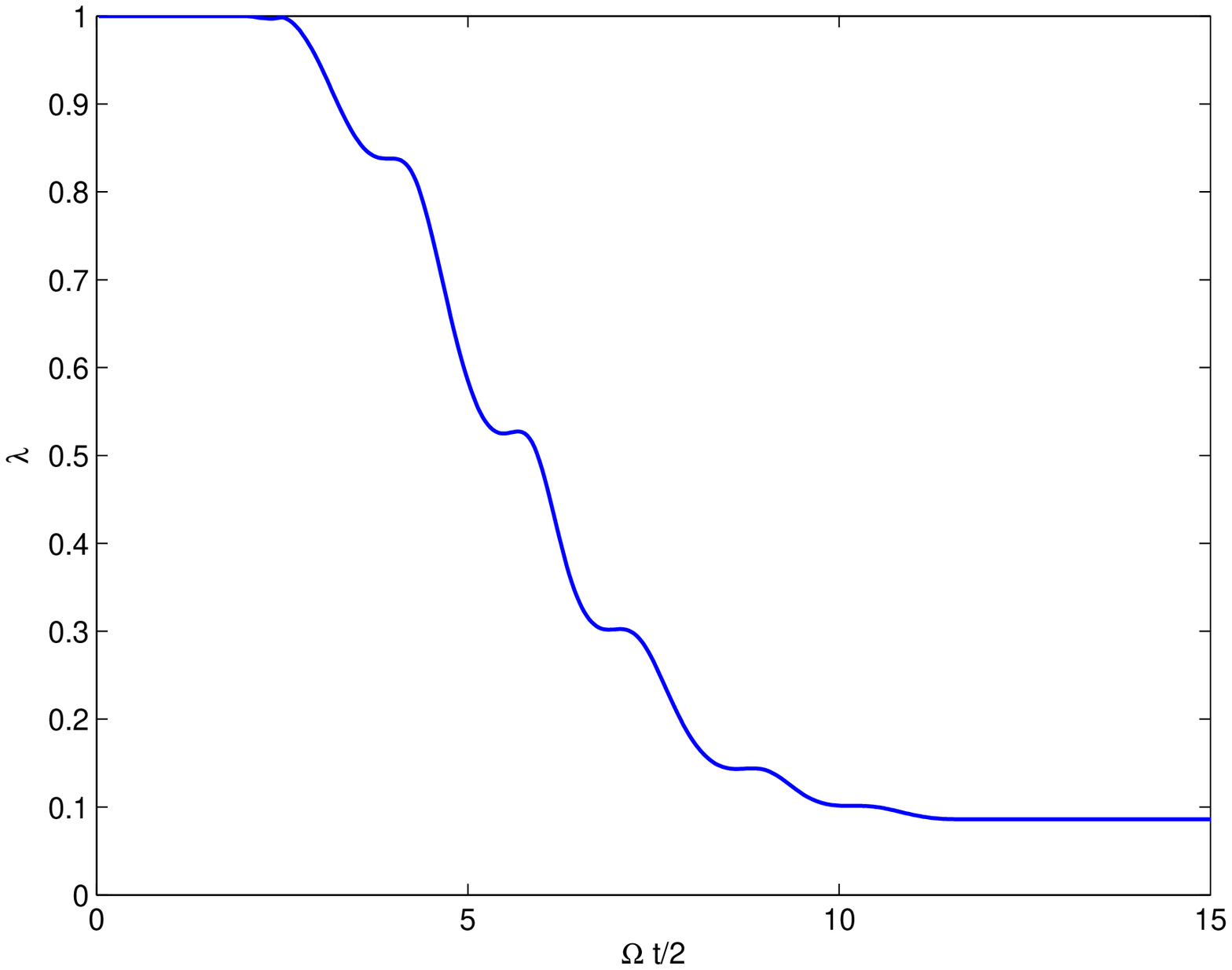}
\includegraphics[scale=0.3]{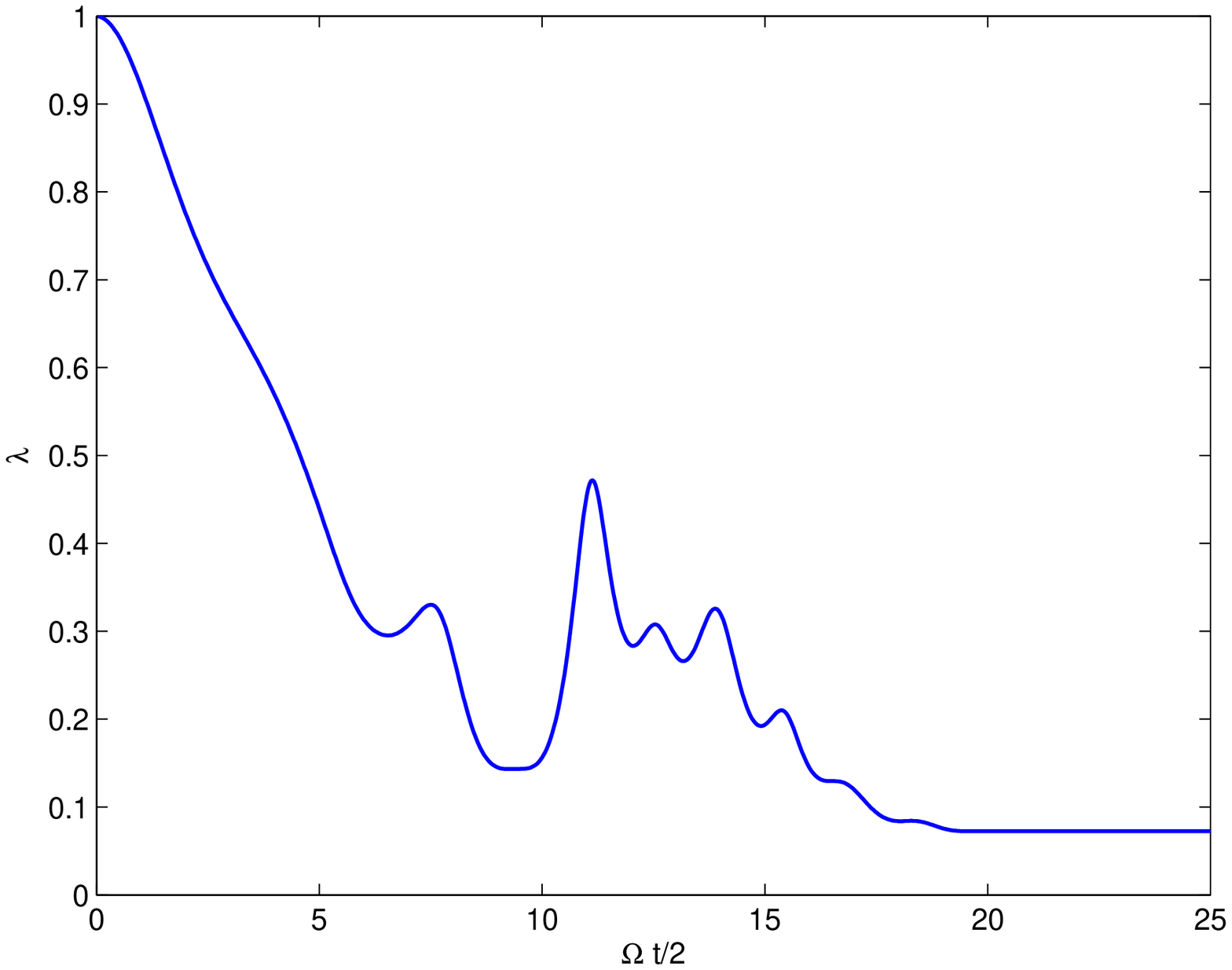}
\caption{Squeezing (smallest eigenvalue of the CM) for an initial state $\sig_0$
(ground state of $\hat{H}_{pw}$) evolving
under the Hamiltonian (\ref{single}) for varying $a$'s and $q$'s.
Left plot: $q(t)=0$ for $0\le t \le 4/\Omega$, $q(t) = 0.5(\Omega t/10-2/5)$
for $4/\Omega\le t \le 14/\Omega$,
$q(t) = 0.5-(0.5)(\Omega t/10-7/5)$ for $14/\Omega \le t \le 24/\Omega$,
$q(t)=0$ for $t\ge 24/\Omega$, and $a(t)=1$ $\,\forall t$.
Right plot: $q(t)=0$ for $0\le t \le 10/\Omega$, $q(t) = 0.5(\Omega t/10-1)$
for $10/\Omega\le t \le 20/\Omega$,
$q(t) = 0.5$ for $20/\Omega\le t \le 30/\Omega$,
$q(t) = 0.5-(0.5)(\Omega t/10-3)$ for $30/\Omega \le t \le 40/\Omega$,
$q(t)=0$ for $t\ge 40/\Omega$,
and
$a(t)=1 - 0.9\Omega t /10$ for $0\le t \le 10/\Omega$, $a(t) = 0.1$
for $10/\Omega\le t \le 20/\Omega$,
$a(t) = 0.1 + 0.9(\Omega t/10 -2)$ for $20/\Omega\le t \le 30/\Omega$,
$a(t) = 1$ for $t\ge 30/\Omega$. \label{Sq2}}
\end{center}
\end{figure}

The left plot of Fig.~\ref{Sq2}
illustrates the effect of a change in the AC component with a steady DC trapping component.
In the case portrayed $a(t)=1$ at all times, while $q$ starts from $q(0)=0$ and increases linearly
up to $0.5$ over a time interval $10/\Omega$ (after a time $4/\Omega$ where the system is kept at
the initial potential and does not evolve). The parameter $q$ then goes back to $0$ over the same
time interval. Finally, the system evolves for another interval lasting $10/\Omega$ under the initial static
trapping Hamiltonian. Even in this case, a degree of squeezing of the $10\%$ ($\lambda=1$) can
be achieved, by changing only the AC potential.
Finally, in the right plot, we report a case where both $a$ and $q$ vary (see the caption for details).
Let us just point out that, as apparent from the plot, a change in both $a$ and $q$ does not in general
grant significant advantages over the, arguably more practical, changes in only one of the parameters.


\begin{figure}[t!]
\begin{center}
\includegraphics[scale=0.3]{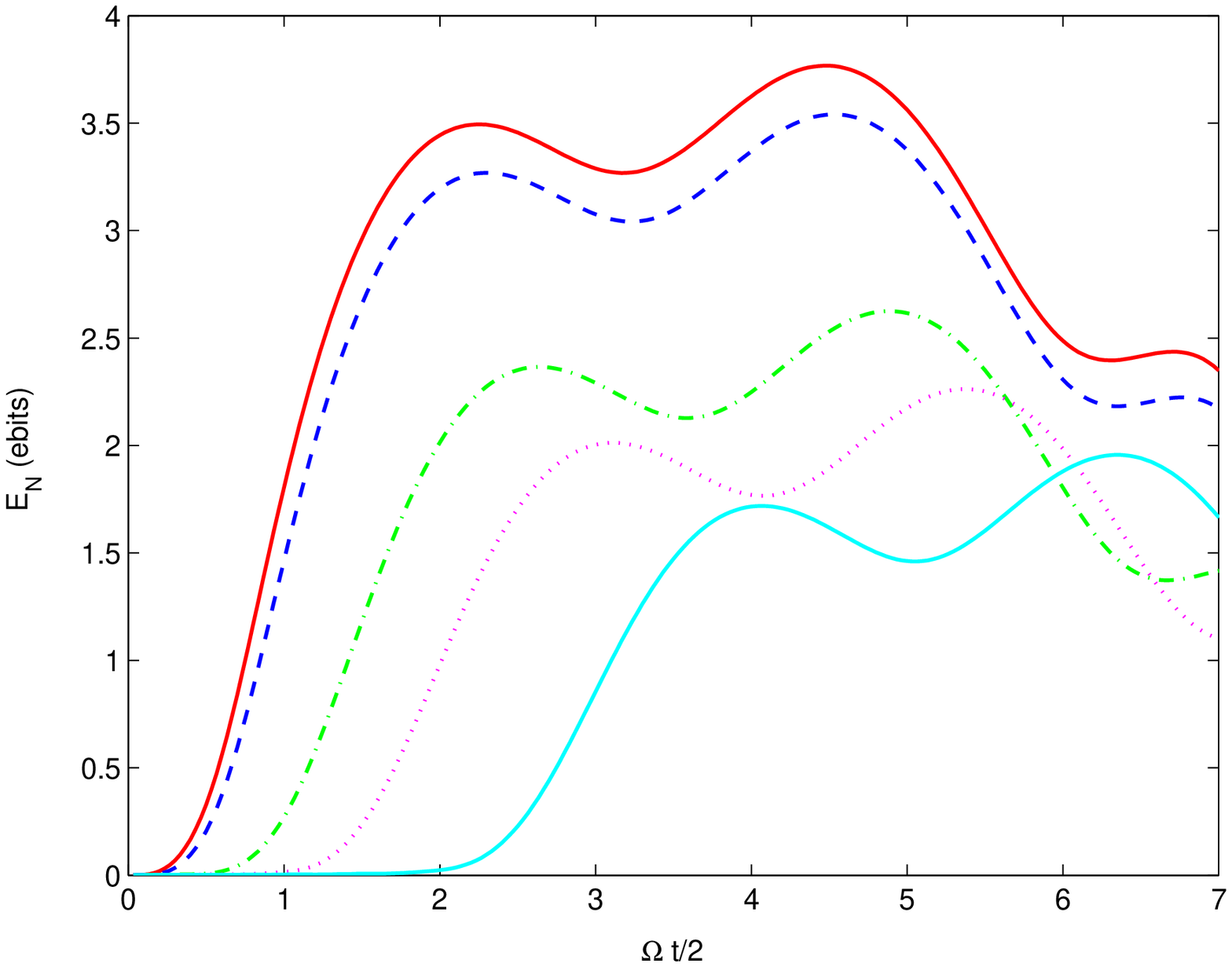}
\includegraphics[scale=0.3]{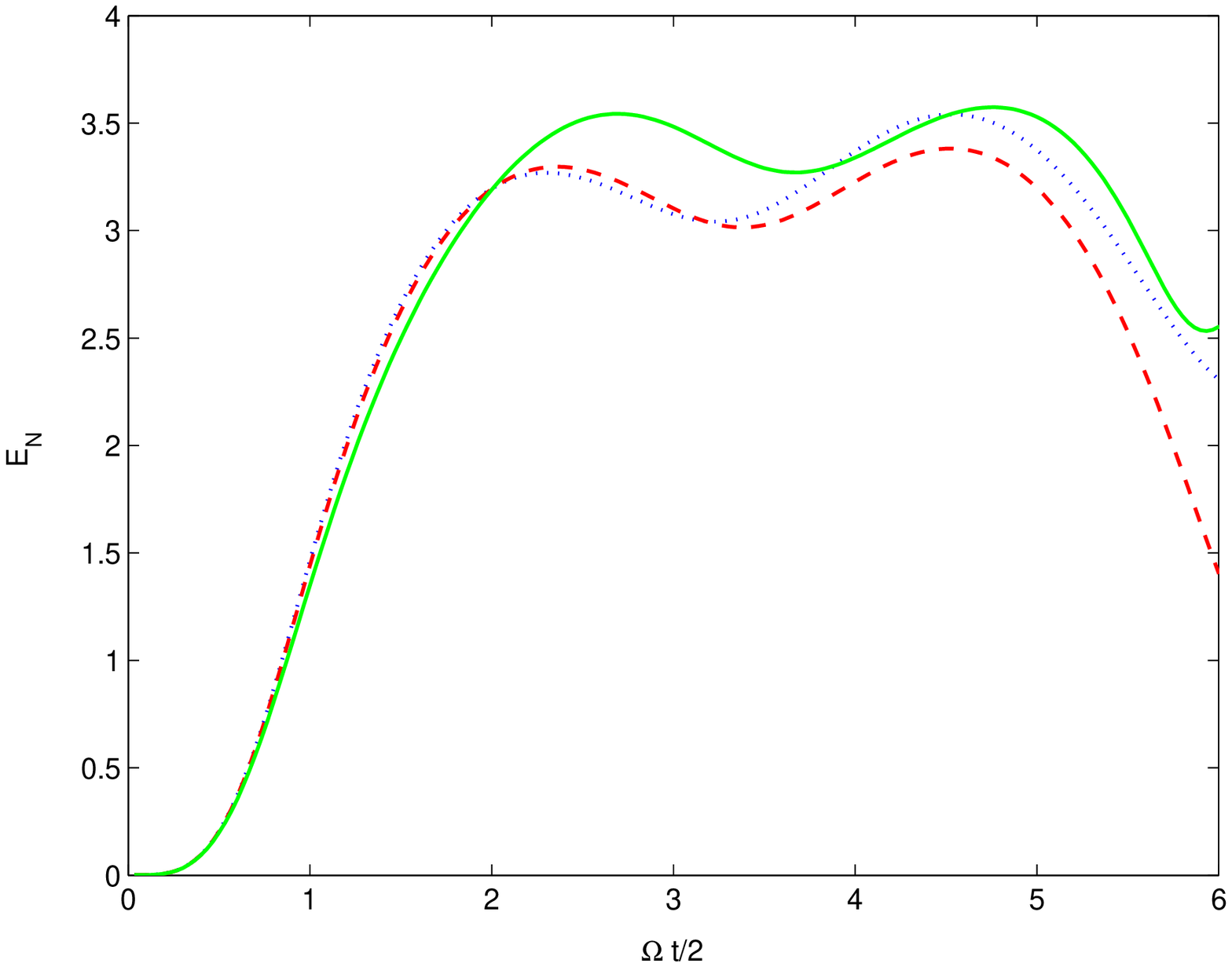}
\caption{Entanglement (logarithmic negativity in ebits) for an initial state $\sig_0$
(ground state of $\hat{H}_{pw}$ for two ions) evolving
under the Hamiltonian (\ref{couple}) for varying $a$'s, different switching rates
and additional oscillating potentials.\newline
On the left hand side, $q(t)=0$ for all the curves.
The system starts off from the ground state for $a(0)=200$ and then
switches linearly in time from $a=200$ to $a=2$, over different time intervals $\Delta t$.
Red (solid) curve: the switching is instantaneous ($\Delta t=0$).
Blue (dashed) curve: $\Delta t=0.2/\Omega$.
Green (dot-dashed) curve: $\Delta t=1/\Omega$.
Magenta (dotted) curve: $\Delta t=2/\Omega$.
Cyan (solid) curve: $\Delta t=4/\Omega$.\newline
On the right hand side, the case of $\Delta t = 0.1$ is portrayed again (blue, dotted),
along with the same case but for $q(t)=0.1$ $\, \forall t$ (red, dashed) and $q(t)=0.5$ $\, \forall t$.
\label{Switch}}
\end{center}
\end{figure}

\section{Generation of entanglement}\label{entanglo}

We shall now consider two ions in a trap and address the evolution of the EPR-like entanglement
between their canonical operators, under the dynamical conditions specified above.
Because the state of the system is Gaussian at all times,
we can quantify such an entanglement by evaluating
the logarithmic negativity, a widely used entanglement monotone related to the absolute
sum of the negative eigenvalues of the partial transposition 
of a quantum state \cite{neg1,neg2,neg3,neg4}.
If $\tilde{\varrho}$ is the partial transposition of the bipartite state $\varrho$
(transposition with respect to only one the two parties' Hilbert spaces), then
the logarithmic negativity $E_N$ of $\varrho$ is given by $E_N = \log_{2} \|\tilde{\varrho}\|_1$,
where $\|\cdot\|_1$ stands for the trace norm.
The logarithmic negativity is an upper bound to the distillable entanglement and
is customarily expressed in ${\rm ebits}$.
It can be
computed for Gaussian states with standard techniques, essentially because the effect of the
partial transposition on positions and momenta is promptly described, and because
such a transformation maps bosonic Gaussian states into bosonic Gaussian states \cite{serafozzi04}.

As in the case of squeezing, we will always start from a ground state of the Hamiltonian
with effective radial trapping frequency $\omega_{pw}$, and where the
corrections due to the Coulomb repulsion are also taken into account.
In the case of two ions, the Coulomb interactions and modifications to the local trapping frequencies
render the study at hand slightly more delicate. This is essentially because instabilities can arise
not only from the configuration of DC and AC trapping fields, but also because of the repulsion
between the ions.
However, the idea behind the generation of entanglement is analogous to that
underlying the generation of squeezing: for large enough initial trapping frequencies,
the ions will start the evolution in a very weakly entangled state
(often separable to most practical effects). If the initial Hamiltonian does not change, the entanglement
will clearly not change either (it will at most oscillate around an average value
if the amplitude of the AC component is large). However, if the parameters of the potential, and
hence the trapping frequencies, change,
the system will perceive such a change as a ``deformation'' of the canonical coordinates, which are
rescaled by the frequencies, that is, essentially, as a squeezing transformation. If, like in the case
of two ions, an interaction term is also present, the squeezing will gradually
be transferred from local coordinates to a non-local combination of the coordinates.
such a squeezing in combined quadratures
corresponds essentially to entanglement in continuous variable systems.

Let us first discuss the role of switching times in the entanglement generation.
The example on the left of Fig.~\ref{Switch} is extremely clear in this respect: only a static trapping field
is considered ($q(t)=0$), with $a(t)$ decreasing from $200$ to $2$
(notice that this correspond to a change of a factor $10$ in the trapping potential,
since $a$ corresponds to a squared trapping frequency) in a time interval $\Delta t$ which
varies from curve to curve, from instantaneous to $\Delta t=4/\Omega$ (from top to bottom).
It is apparent that faster switching rates allow for a superior entanglement generation.
In fact, while the Hamiltonian is changing, the ground state of the system ``adapts''
to the new Hamiltonian if the change is too slow (much in the spirit of the adiabatic theorem).
The actual entanglement generation only begins once the trapping potential 
reaches the new value, and its magnitude
will depend on the rapidity of the change.
In general, switching rates of the order of $10\sqrt{a}\Omega$ 
allow for a close to ideal creation of entanglement, but substantial entanglement is also there for 
switching slower by one order of magnitude.

The right side of Fig.~\ref{Switch} shows the effect of an added AC component on the
same evolution, for $\Delta t=0.1$. As evident from the plot, an AC with $q$ up to $0.5$
affects only rather marginally the evolution of the logarithmic negativity. In general,
moreover, the effect on the entanglement of additional oscillating potentials is erratic
and does not monotonically depend on the AC amplitude.

\begin{figure}[t!]
\begin{center}
\includegraphics[scale=0.4]{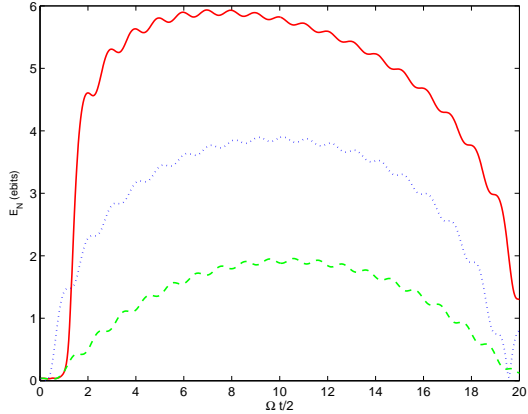}
\caption{Entanglement (logarithmic negativity in ebits) for an initial state $\sig_0$
(ground state of $\hat{H}_{pw}$ for two ions) evolving
under the Hamiltonian (\ref{couple}) for varying $q$'s and different switching rates.
$a(t)=10$ for all curves.
In all cases, the system starts from the ground state for $a(0)=10$ (and $q(0)=0$), and then
switches linearly in time from $q=0$ to $q=100$, over different time intervals $\delta t$'s.
Red (solid) curve: $\delta t=1.3/\Omega$).
Blue (dotted) curve: $\Delta t=0.2/\Omega$.
Green (dashed) curve: $\Delta t=1/\Omega$.\label{AC}}
\end{center}
\end{figure}

Similar perfomances can be obtained by keeping the same static potential and varying
the AC component. Fig.~\ref{AC} shows a non-trivial instance of such dynamics.
The system starts from the ground state for $a=10$ and $q=0$. Then,
the parameter $q$ controlling the AC amplitude is linearly increased to $100$ and
turned off again over
different time intervals $2\delta t$'s. The middle curve (blue, dotted) refers
to $\delta t = 0.2 / \Omega$: in this case the switch is rather fast and the entanglement generation
substantial.
The lower curve (green, dashed) refers to $\delta t = 1 / \Omega$: the switching is slower
and thus less entanglement is created.
The upper curve (red, solid) refers to $\delta t = 1.3 / \Omega$: here the entanglement
is larger than in the previous case. In fact, the transient dynamics of the system mostly takes place
in a region of parameters which is definitely unstable: as we have seen in the case of squeezing
generation, spending a sizeable part of the dynamics in such regions can create
very high squeezing and hence, in this case, entanglement. This curve shows a sudden boost
in entanglement right after the transient interval, which is a signature of `impending' instability:
in fact, higher $\delta t$'s would be impractical, because the ion would probably get lost (the numerics start
diverging there). It shouldn't surprise that the entanglement keeps oscillating over large time-scales
after the initial Hamiltonian is re-established. This is due to the fact that, once the squeezing is
generated through the varying potentials, the Coulomb interaction keeps rotating the state 
in phase space, making it undergo cycles of entanglement and disentanglement.

\section{Conclusions and outlook}\label{outro}

Summing up, in the present note we have shown that:
\begin{itemize}
\item{both entanglement and squeezing of the motional degrees of freedom of
trapped ions can be effectively created by controlling the AC and/or the DC
component of the trapping potentials;}
\item{regions of trapping instability can be profitable
to boost the generation of motional squeezing and entanglement (if the permanence
in such regions is short enough not to lose the ions!);}
\item{switching rates of the order of $10\sqrt{a}\Omega$, that is of the order of the effective 
trapping frequency, are ideal to generate such resources
(although one order of magnitude less still yields interestingly good values).}
\end{itemize}
Note that, in principle, the squeezing and entangling operations presented here could be iterated
to achieve muchly improved performances (see \cite{citeus} for details), the ultimate limit
being essentially the tolerance of the trap's geometry to large displacements
(large squeezing in positions and momenta implies in fact broad oscillations).

Let us also remark that operations on the ions can be realised also by controlling 
the RF frequency $\Omega$, while leaving the strenghts of the potentials unchanged.
This path has not been followed in the present paper.

To conclude, let us point out that, in view of the considerable potential demonstrated
in the present and previous investigations and of the widespread interest in
generating and distributing optical squeezing and entanglement, one could argue that the ultimate
applicability of this sort of manipulations should be aimed at hybrid systems where,
after the resources are generated {\em in situ} by controlling the potentials,
the ions are then coupled to light through cavities and the squeezing or entanglement are 
swapped to optical modes.
Future work will focus on this possibility \cite{futureus}.

Finally, a further line of investigation could focus on the possibility 
of realising any quantum gate (not restricted to Gaussian operations) 
between the motions of two ions, by exploiting 
anharmonicities and the control of the potentials' parameters.

\end{document}